\documentclass[twocolumn,aps,prl,lengthcheck,noshowpacs]{revtex4}
\usepackage[dvips]{graphicx}
\usepackage{amssymb}
\usepackage{amsmath}
\usepackage{latexsym}
\usepackage{epsfig}
\usepackage{bm}
\usepackage{color}
\usepackage{sistyle}
\usepackage{times,psfrag,subfigure}      
\begin{document}

\title{Dynamic Morphologies and Stability of Droplet Interface Bilayers} 
\author{Benjamin Guiselin$^\dagger$, Jack O. Law$^\dagger$, Buddhapriya Chakrabarti$^\ddagger$ and Halim Kusumaatmaja$^\dagger$} 
\email{halim.kusumaatmaja@durham.ac.uk}
\affiliation{$^\dagger$Department of Physics, Durham University, Durham DH1 3LE, U.K. \\ $^{\ddagger}$Department of Physics and Astronomy, University of Sheffield, Sheffield S3 7RH, U.K.}
\date{\today}

\begin{abstract}
We develop a theoretical framework for understanding dynamic morphologies and stability of droplet interface bilayers (DIBs), accounting for lipid kinetics in the monolayers and bilayer, and droplet evaporation due to imbalance between osmotic and Laplace pressures. Our theory quantitatively describes distinct pathways observed in experiments when DIBs become unstable. We find that when the timescale for lipid desorption is slow compared to droplet evaporation, the lipid bilayer will grow and the droplets approach a hemispherical shape. In contrast, when lipid desorption is fast, the bilayer area will shrink and the droplets eventually detach. Our model also suggests there is a critical size below which DIBs can become unstable, which may explain experimental difficulties in miniaturising the DIB platform. 
\end{abstract}

\maketitle

\begin{figure}[t]
\centering
\includegraphics[width=0.9\linewidth]{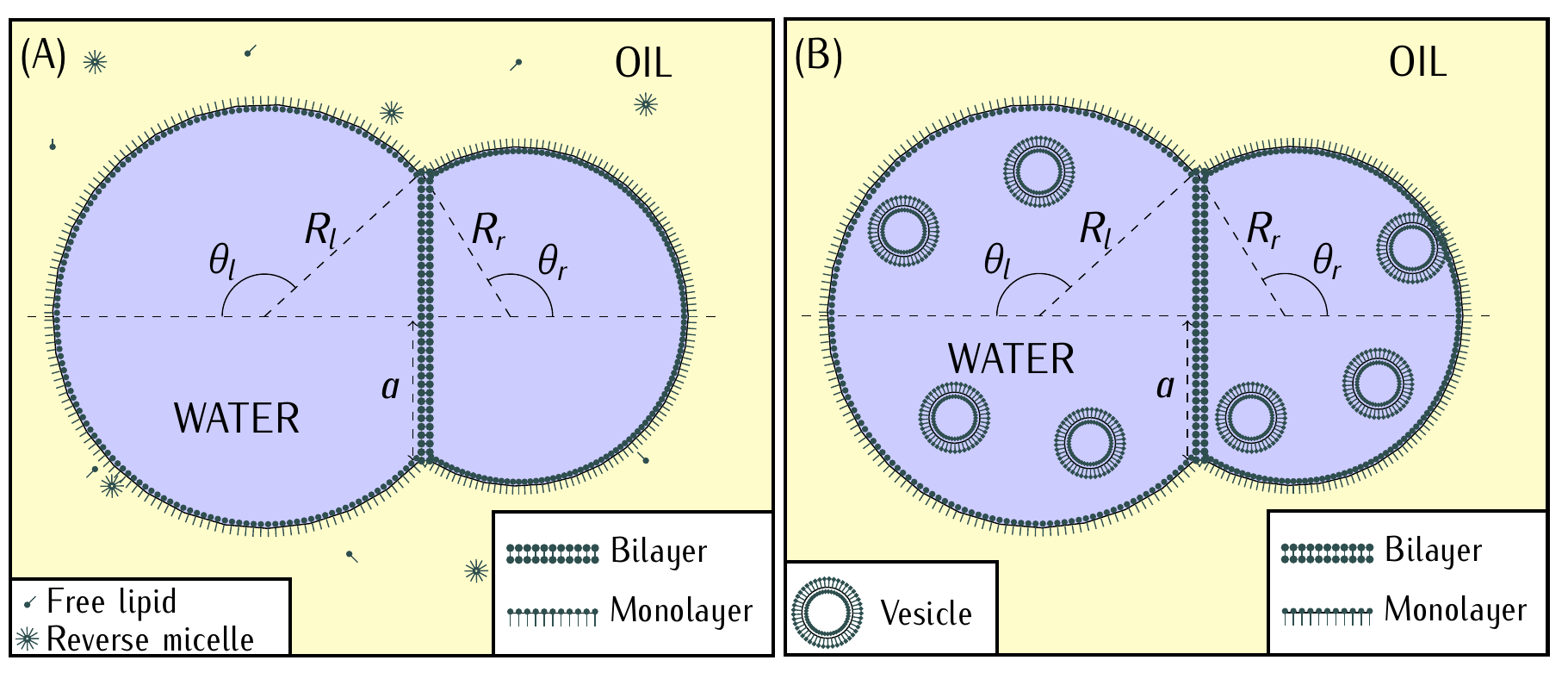}
\caption{Schematic diagrams of (a) lipid-out and (b) lipid-in Droplet Interface Bilayers (DIBs).}
\label{fig:schematic}
\end{figure}

Droplet Interface Bilayers (DIBs) are constructed by bringing together two (or more) lipid monolayer-encased water droplets submerged in oil \cite{Funakoshi2006,Bayley2008DIB}. As the droplets contact one another, lipid bilayers form spontaneously. The lipids can be introduced in the bulk of the oil phase (lipid--out, Fig. \ref{fig:schematic}(a)) or inside the water droplets (lipid--in, Fig. \ref{fig:schematic}(b)). DIBs can be assembled in several ways, including connecting millimeter-sized aqueous droplets using pipettes, electrodes or lasers \cite{Funakoshi2006,Bayley2008DIB,Leptihn2013,Dixit2012}, high throughput microfluidic devices \cite{Elani2012,Thutupalli2013Micro,Thiam2012,Czekalska_micro} and 3D printing \cite{Villar2014Tissue}. 

DIBs have a number of advantages over other lipid bilayer platforms. Electrical characterisation across the bilayer is easy to perform \cite{Hwang2007Electrical,Poulos2009,Freeman2015}. It is possible to introduce asymmetric bilayers using the lipid-in method \cite{Hwang2008Asymmetric,Milianta2015} and to construct complex droplet networks \cite{Elani2013,Schlicht2015,Holden2007}. A number of membrane proteins have also been successfully reconstituted across DIBs, including the viral potassium channel Kcv, the light-driven proton pump bacteriorhodopsin, and the mechanosensitive channel of large conductance (MscL) \cite{Holden2007,Barriga2014,Syeda2008,Poulos2009Ion}. Given these advantages, the potential applications of DIBs are wide-ranging, from droplet arrays for ion channel screening \cite{Syeda2008,Poulos2009Ion} and chemical microreactors \cite{Elani2014,Elani2016Microfluidic} to responsive materials  \cite{Punnamaraju2012,Villar2014Tissue,Zhang2016Shape} and mimics of electrical circuits and logic gates \cite{Maglia2009Droplet}. 

The stability of DIBs, however, remains a major issue \cite{Leptihn2013,boreyko2013evaporation,mruetusatorn2014dynamic}, especially for DIBs below several hundreds of microns \cite{boreyko2013evaporation}. Furthermore, when DIBs become unstable, their morphological evolution is extremely rich \cite{mruetusatorn2014dynamic}. As the droplets shrink due to evaporation, the lipid bilayer can (i) zip and increase in size, (ii) unzip until the droplets eventually detach, or (iii) the system can shrink almost uniformly. Here we provide a theoretical framework to address both the issues of DIB stability and dynamic morphologies, for which there has been no explanation to date. The ingredients of our model are the balance between Laplace and osmotic pressures, which determine the evaporation rate of DIBs; and lipid kinetics, which include lipid adsorption, desorption and exchange between the mono- and bi-layers. 


Our main results are as follows.~First, we are able to construct comprehensive phase diagrams, which  reproduce and distinguish the dynamic regimes observed in experiments. Second, our theory predicts a size limit of stable DIBs. For the typical materials used in references \cite{boreyko2013evaporation,mruetusatorn2014dynamic}, DIBs smaller than O(100 $\mu$m) can become unstable. Third, we elucidate a mechanistic understanding for both the dynamic morphology diagram and DIB stability, arising out of a competition between four characteristic timescales: lipid desorptions for the (i) mono- and (ii) bi-layers, (iii) droplet evaporation, and (iv) lipid exchange between the mono- and bi-layers.

We begin by describing our model for lipid kinetics. We use subscripts $i=l,r$ to represent the left and right droplets (see Fig. 1), superscript $b$ to label the bilayer, and no superscript for the monolayers (for brevity). The change in the number of lipids on the monolayers and bilayer is due to three different processes. The first process is lipid adsorption from the bulk liquid to the interface, 
\begin{equation}
\frac{d}{dt}\left(\Gamma_iS_i\right)_{ad}=k_{on}\left(\Gamma_{\infty}-\Gamma_i\right)S_i. \label{adsorption}
\end{equation}
The adsorption rate is proportional to the density of available sites per unit area, $\left(\Gamma_{\infty}-\Gamma_i\right)$. $\Gamma_{\infty}$ is the density of total available sites, while $\Gamma$ is the density of occupied sites. $k_{on}$ is the rate constant of lipid adsorption, and $S$ is the monolayer area. 

The second process is lipid desorption from the interface to the bulk liquid, 
\begin{equation}
\frac{d}{dt}\left(\Gamma_iS_i\right)_{de}=k_{off}\Gamma_iS_i, \label{desorption} 
\end{equation}
with $k_{off}$ the desorption rate constant. In general $k_{on}$ and $k_{off}$ depend on the lipid density in the bulk liquid \cite{eastoe2000dynamic}. There are several possible underlying molecular mechanisms for lipid transfer, including rupture and extraction mechanisms \cite{venkatesan2015adsorption}. However, in our minimal model, it is neither necessary nor possible to make explicit statement about molecular mechanisms. For simplicity, we assume $k_{on}$ and $k_{off}$ to be constants.

The third process is the exchange of lipids from the monolayer to the bilayer, and vice versa,
\begin{equation}
\frac{d}{dt}\left(\Gamma_iS_i\right)_{mb}=\frac{2\pi a \xi}{k_B T} \left[ \mu^{b}(\Gamma^{b}) - \mu_i(\Gamma_i) \right]. \label{current_monolayer_bilayer}
\end{equation}
We assume the current to be proportional to the difference in chemical potentials for the lipids in the monolayers and bilayer, with a proportionality constant $\xi$. The prefactor $2\pi a$ corresponds to the contact line perimeter where the monolayers meet the bilayer, with $a$ the bilayer radius and $A=\pi a^2$ the bilayer area. Here $k_B$ is the Boltzmann constant and $T$ is the system temperature.

The chemical potential of the monolayer can be calculated using the Gibbs-Duhem equation \cite{Doi2013}, $-\Gamma d\mu=d\gamma$. We employ a standard relation between the monolayer tension and its surface excess \cite{liu2000diffusion}, $\gamma_i(\Gamma_i)=\gamma_{0}+k_{B}T \, \Gamma_{\infty}\ln \left(1-\Gamma_i/\Gamma_{\infty}\right)$. $\gamma_{0}$ is the surface tension of a clean oil-water interface, when there is no adsorbed lipid. Substituting this equation to the Gibbs-Duhem equation leads to the following relation for the chemical potential
\begin{equation}
\mu_i = k_B T \ln\left[\Gamma_i / (\Gamma_{\infty} - \Gamma_i)\right].
\end{equation} 

Similar to Eqs. \eqref{adsorption}--\eqref{current_monolayer_bilayer}, equivalent relations for lipid adsorption, desorption and exchange can be written for the bilayer, as given in the SI \cite{supplementary}. For the lipid-out method, there is no direct lipid exchange from the bulk liquid to the bilayer. The variation in the number of lipids is only due to the exchange between the monolayers and bilayer. Additionally, we assume the bilayer to be incompressible with constant chemical potential $\mu^{b}$ and lipid density $\Gamma^b$. Changes in the total number of lipids in the bilayer thus 
necessarily involve changes in the bilayer area. To justify this assumption, we note that bending deformation to the bilayer is not appreciable in the typical experiments \cite{boreyko2013evaporation,mruetusatorn2014dynamic}. Using the DOPC bilayer as an example, the compressibility is $\approx \SI{290}{mN.m^{-1}}$ while its bending modulus is $\approx 29\ k_{B}T$ \cite{venable2015mechanical}. Consequently, for an energy scale of order of the bending energy, the relative change in lipid density on the bilayer due to compression is negligible, of order $0.01\%$. Indeed for higher compression, the lipid bilayer buckles \cite{boreyko2013evaporation,mruetusatorn2014dynamic}.

To model droplet evaporation, here we focus on cases where the two droplets forming the DIBs are exactly or close to being symmetric. Thus, evaporation is driven by the imbalance between the osmotic pressure difference $\Delta \Pi$ and the Laplace pressure $P^{L}$ between outside and inside the droplets. The osmotic pressure difference and the Laplace pressure between the two droplets themselves are negligible. The outward flux of water from droplet $i$ can be written as \cite{staykova2013confined}
 \begin{equation}
J_i^{out}= -\frac{dV_i}{dt}=\frac{p_{f}v_{w}}{R_{GP}T}\left(\Delta \Pi_i-P^{L}_i\right)S_i.
\label{water_flow}
\end{equation}
$v_{w}$ is water's molar volume, $R_{GP}$ is the gas constant, and $p_{f}$ is the monolayer's water permeability coefficient. The osmotic pressure outside the droplets, $\Pi^{out}$, is taken to be constant with time, while inside the droplets we use van't Hoff's law \cite{Doi2013} for dilute solutions, $\Pi^{in}_i=C^{in}_i V_i\left(0\right) R_{GP} T / V_i$, with $C_i^{in}$ the inside osmolarity and $V_i\left(0\right)$ the initial droplet volume. The Laplace pressure is $P^{L}_i=\ 2\gamma_i(\Gamma_i) / R_i$, where $R_i$ is the radius of droplet $i$, see Fig. \ref{fig:schematic}.

\begin{figure*}[t]
\centering
\includegraphics[width=0.87\linewidth]{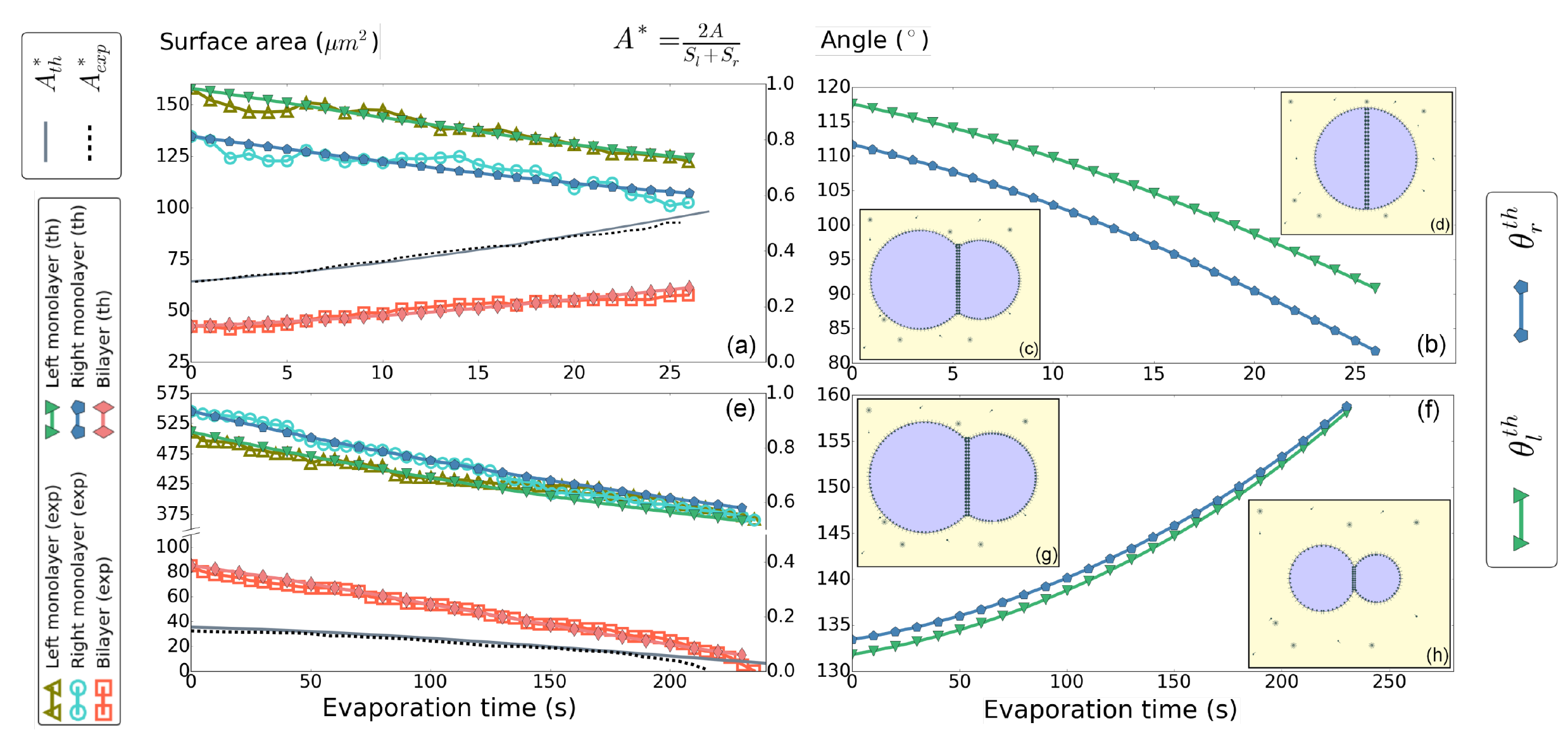}
\caption{(a and e) Surface areas and bilayer area fraction as a function of time superposed with experimental data \cite{mruetusatorn2014dynamic} for the ``bilayer expansion'' and ``unzipping'' modes, where the bilayer area expands and shrinks respectively. (b and f) In the ``bilayer expansion'' (resp. ``unzipping'') mode, this is accompanied by a decrease (resp. increase) in the polar angle. The insets show the initial DIB configurations (c and g) and their configurations upon evaporation (d and h). The best fit parameters to the experimental data are: 
$C^{in}_{l/r}=(8.58\pm 0.07) \times \SI{e-2}{mol.m^{-3}}$, $\Pi^{out}=(4.34 \pm 0.03) \times \SI{e4}{Pa}$, $k_{off}=(4.78 \pm 0.04) \times \SI{e-3}{s^{-1}}$, $\Gamma(0)=\SI{1.42e18}{m^{-2}}$ and $\mu^{b}=(7.43\pm 0.02) \times \SI{e-2}{} k_BT$ for panels (a-d); and $C^{in}_{l}=(1.74 \pm 0.01) \times \SI{e-2}{mol.m^{-3}}$, $C^{in}_{r}=(0.72 \pm 0.01) \times \SI{e-2}{mol.m^{-3}}$, $\Pi^{out}=(2.19 \pm 0.03) \times \SI{e4}{Pa}$, $k_{off}=(2.50\pm 0.01) \times \SI{e-3}{s^{-1}}$, $\Gamma(0)=(1.620\pm 0.004) \times \SI{e18}{m^{-2}}$ and $\mu^{b}=(9.96\pm 0.07) \times \SI{e-1}{}k_BT$ for panels (e-h). In all cases we have used $\xi=(9.85 \pm 0.34) \times \SI{e10}{(m.s)^{-1}}$.
}
\label{fig:fitting}
\end{figure*}

Recent experiments demonstrate that as DIBs become unstable, their dynamic morphologies can follow several distinct pathways. Mruetusatorn {\it et al.} \cite{mruetusatorn2014dynamic} reported three classes of behaviour. Class I (``bilayer expansion'') is highlighted by the observation that the bilayer area grows (Fig. \ref{fig:fitting}(a)) and the droplets' polar angles decrease (Fig. \ref{fig:fitting}(b)) upon evaporation. The polar angles, $\theta_l$ and $\theta_r$, are defined in Fig.~\ref{fig:schematic}. In contrast, for class III (``unzipping''), the bilayer area shrinks (Fig. \ref{fig:fitting}(e)) while the polar angles increase (Fig. \ref{fig:fitting}(f)) until eventually the two droplets detach. Finally, in class II, the DIB shrinks while maintaining approximately constant polar angles (see SI \cite{supplementary}, including comparison with our model). According to our theory, class II corresponds to one of several possible crossover behaviours between the two dominant dynamic modes: ``bilayer expansion''  and ``unzipping''.

Fig. \ref{fig:fitting} demonstrates that our model reproduce experimental results with good agreement. The experimental data were obtained using the lipid-out approach, in which  DOPC lipids are introduced in soybean oil (top row) and hexadecane (bottom row). As the droplet evaporates, we find the lipid surface density $\Gamma$ closely approaches the saturated density $\Gamma_\infty$. Thus, far from equilibrium, lipid adsorption at the monolayer is very small compared to desorption, $k_{on}\left(\Gamma_{\infty}-\Gamma_i\right) \ll k_{off}\Gamma_i$, and can be neglected. To reduce the number of free parameters, we assume the initial lipid monolayer density is the same for the right and left droplets, and equal to the bilayer lipid density, $\Gamma^b=\Gamma(0)$. We also assign typical literature values to the following parameters: $p_{f}=\SI{80}{\micro m.s^{-1}}$ \cite{walter1986permeability,mathai2008structural,milianta2015water,olbrich2000water,finkelstein1976water}, $\Gamma_{\infty}=\SI{2.3e18}{m^{-2}}$ \cite{sehgal2003micellar}, and $\gamma_{0}=\SI{25}{mN.m^{-1}}$ and $\SI{44}{mN.m^{-1}}$ for soybean oil/water and hexadecane/water interfacial tensions \cite{zhou2013facile}. The remaining parameters in our theory ($\xi$, $k_{off}$, $\Gamma(0)$, $\mu^{b}$, $\Pi^{out}$, $C_{l,r}^{in}$) are optimized against experimental data using a Markov Chain Monte Carlo (MCMC) method \footnote{The value of $\xi$ is optimized to fit the experimental data in Fig. \ref{fig:fitting}(a), and its numerical value is kept for other simulations. We only need to fit $\Gamma(0)$ for the hexadecane/water system (Fig. \ref{fig:fitting}(e)), since the availability of experimental surface tension data for the soybean oil/water system allows us to directly determine the initial lipid density.}. The details are presented in SI \cite{supplementary}. The best-fit parameters are given in the caption of Fig. \ref{fig:fitting}. They compare well to literature values for similar systems, tabulated in SI \cite{supplementary}.

\begin{figure}[ht]
\centering
\includegraphics[width=0.99\linewidth]{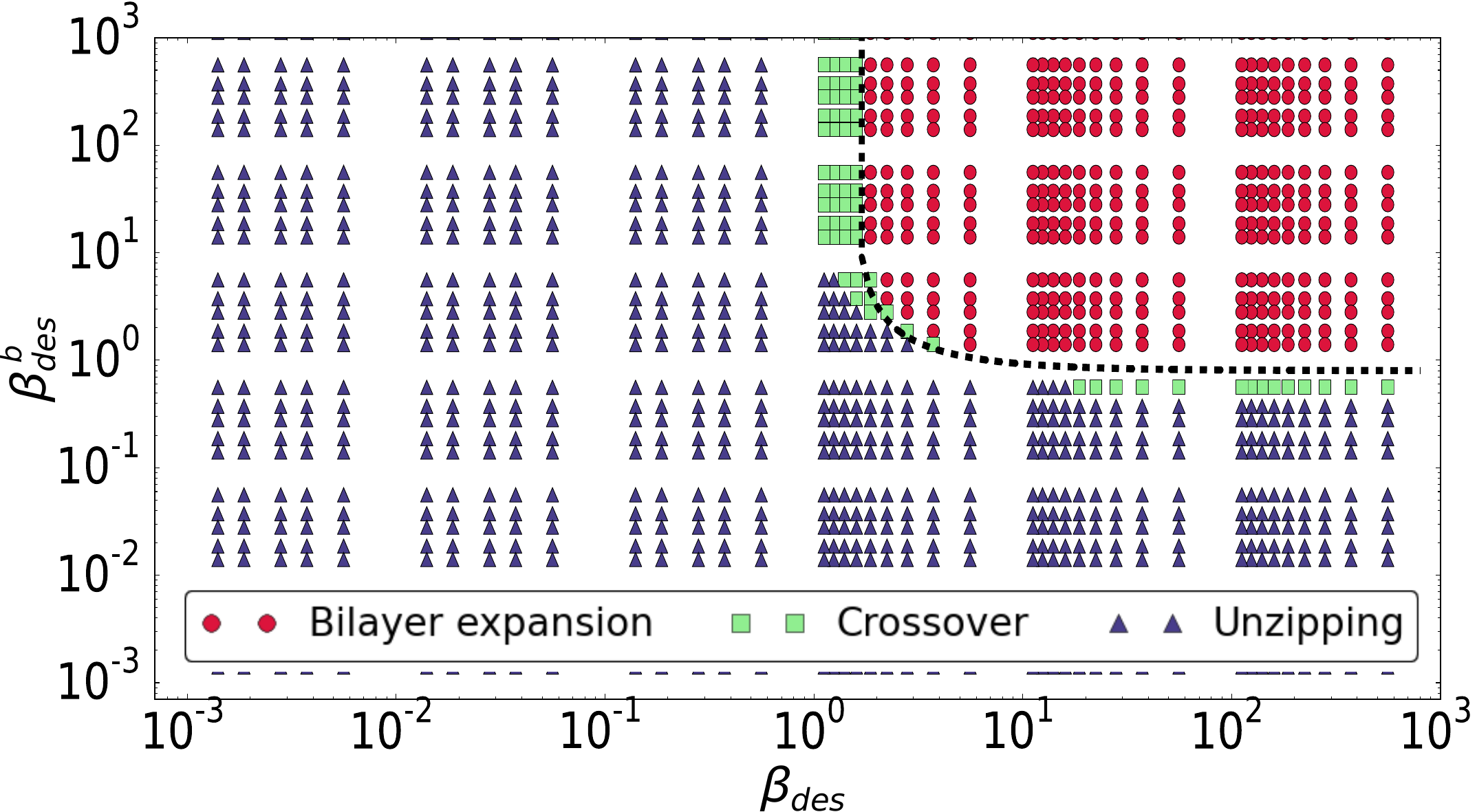}
\caption{Dynamic morphology diagram in terms of timescale ratios between lipid desorptions and droplet evaporation, $\beta_{des}$ and $\beta_{des}^{b}$. The bilayer expansion mode is observed when desorptions are slow, and the unzipping mode when one of the desorption timescales is fast. The dotted line is a guide to the eye separating these two regions, and the crossover behaviours are detailed in SI \cite{supplementary}. We use $\mu^{b}=\SI{0.25}{} k_B T$, $\Pi^{out}=\SI{2.48e4}{Pa}$ and $C_{l/r}^{in}=\SI{5e-2}{mol.m^{-3}}$. The other parameters are as in Fig. \ref{fig:fitting} (top row) for DOPC at soybean oil/water interface.}
\label{fig:desorption}
\end{figure}

Beyond being able to fit the reported experimental data, an important insight from our theory is that we can explain the key factors determining the observed dynamical pathways. From Eqs. \eqref{desorption}, \eqref{current_monolayer_bilayer} and \eqref{water_flow}, we identify four characteristic timescales: (i-ii) lipid desorptions for the mono- and bi-layers, $\tau_{des}=1/k_{off}$ and $\tau^{b}_{des}=1/k^{b}_{off}$; (iii) droplet evaporation,  $\tau_e= (R_{GP}TR)/(p_f v_w \Pi^{out})$; and (iv) lipid exchange from the mono- to bi-layer, $\tau_{flow}=\Gamma_{\infty} R/\xi$. We construct a dynamic morphology diagram in Fig. \ref{fig:desorption}, concentrating on the role of the lipid desorption and droplet evaporation timescales. We vary $\beta_{des} = \tau_{des}/\tau_e$ and $\beta^{b}_{des}=\tau^{b}_{des}/\tau_e$ while keeping $\tau_{flow}/\tau_e=0.59$ constant. We also neglect lipid adsorption, following the results in Fig.  \ref{fig:fitting}, where it is negligible compared to lipid desorption. We observe two dominating regimes. The first one is broadly defined by $\beta_{des}\gtrsim 1$ and $\beta_{des}^{b}\gtrsim 1$: if desorption is slow, to compensate the increase in lipid monolayer density upon evaporation (shrinkage in monolayer area), lipids flow from the monolayer to the bilayer, resulting in bilayer zipping and growth. In contrast, for $\beta_{des}\lesssim 1$ or $\beta_{des}^{b}\lesssim 1$, desorption from either the mono- or bi-layer is fast enough to tackle the rise in lipid density. As the droplets shrink, the bilayer unzips and the droplets detach. 

Between these two dominating behaviours, we observe a crossover regime, which occurs when the evaporation timescale becomes similar to the desorption timescale ($\beta_{des}\approx 1$ or $\beta_{des}^{b}\approx 1$). Class II behaviour reported by Mruetusatorn {\it et al.} \cite{mruetusatorn2014dynamic} is one of three possible crossover behaviours between the ``bilayer expansion'' and  ``unzipping'' modes (see SI \cite{supplementary}). Varying the value of $\tau_{flow}/\tau_e$ leads to similar dynamic morphology diagram, where the ``bilayer expansion'' mode is favoured at large $\beta_{des}$ and $\beta_{des}^{b}$, while the ``unzipping'' mode is favoured for small $\beta_{des}$ or $\beta_{des}^{b}$. With decreasing $\tau_{flow}/\tau_e$, the boundary between the two dominant modes shifts to larger $\beta_{des}$ and smaller $\beta_{des}^{b}$ \footnote{Consider the case where lipid desorption is faster in the bilayer compared to the monolayers. When lipids can exchange  easily from the monolayers to the bilayer, the ``bilayer expansion'' mode is favoured. For the case where lipid desorption is faster in the monolayers compared to the bilayer, flow of lipids from the bilayer to the monolayers will result in bilayer ``unzipping''.}.

We now focus on the equilibrium states of DIBs and their relative stability.  For simplicity, the left and right droplets are taken to be identical. At equilibrium, lipid adsorption is balanced by desorption. Equating Eqs. \eqref{adsorption} and \eqref{desorption} gives equilibrium lipid density, $\Gamma_i=\Gamma_{\infty}k_{on}/\left(k_{on}+k_{off}\right)$. Furthermore, the chemical equilibrium between the mono- and bi-layers imposes the chemical potential of lipids on the bilayer: $\mu^{b}=\mu=k_BT \ln\left(k_{on}/k_{off}\right)$. Finally, from Eq. \eqref{water_flow}, the balance between the Laplace and osmotic pressures gives the equilibrium radius of the droplets:
\begin{equation}
\label{radius_equilibrium}
R^{*}=\frac{2}{\Delta \Pi}\left[\gamma_{0}-k_B T \Gamma_{\infty}\ln\left(1+\frac{k_{on}}{k_{off}}\right)\right].
\end{equation}
Thus, to realise stable DIBs at different sizes for a given set of materials (lipids and oils), it is necessary to adjust the osmotic pressure difference inside and outside the droplets.

\begin{figure}[t]
\centering
\includegraphics[width=0.99\linewidth]{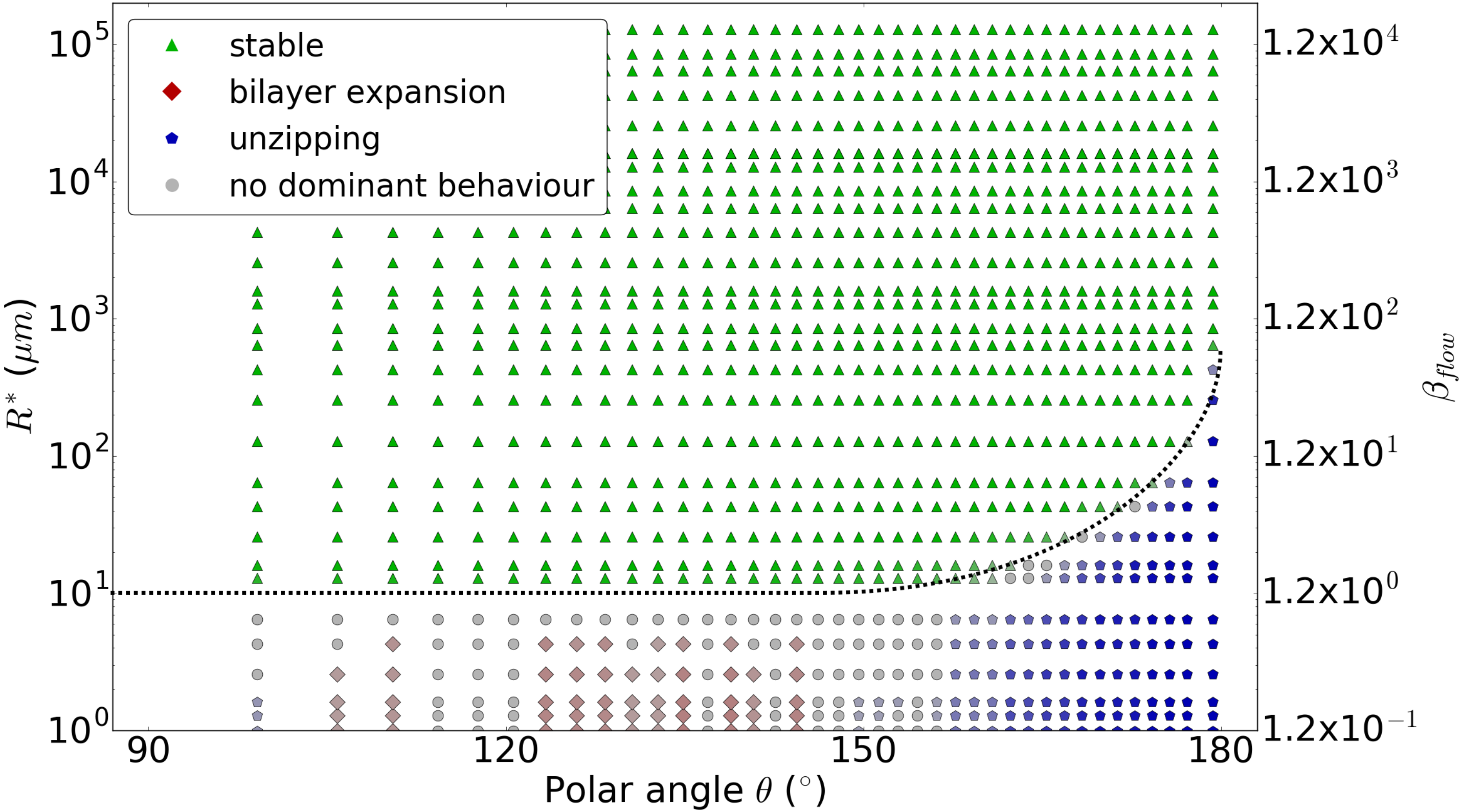}
\caption{Stability diagram of DIBs in terms of the droplet radius $R^{*}$ (equivalently $\beta_{flow}$) and polar angle. The dotted lines are guides to the eye. The colour intensity for each data point signifies the propensity of the simulation outcomes over 120 runs with random perturbations to the equilibrium configuration. Data points with no dominant behaviour (defined as $>55\%$ of the outcomes) are shown in grey circles. Since we start the simulations close to equilibrium, both lipid adsoprtion and desorption are relevant. We use $k_{off}=\SI{5e-3}{s^{-1}}$ and $k_{on}=\SI{8e-3}{s^{-1}}$. The other parameters are as in Fig. \ref{fig:fitting} (top row) for DOPC at soybean oil/water interface.}
\label{fig:stable}
\end{figure}

Next, starting from the equilibrium states, we perturb the DIB morphologies by randomly varying the outside osmotic pressure, bilayer radius, monolayer lipid densities and polar angles of the droplets within $\pm 10\%$ of the initial equilibrium value. In agreement with experimental observations \cite{boreyko2013evaporation}, the DIBs are not always stable. Fig. \ref{fig:stable} shows a transition between stable and unstable DIBs at around 10 $\mu m$ to 1 mm depending on the polar angle. The simulation parameters are provided in the caption. Each data point is the average behaviour from 120 random perturbations \footnote{We have checked that the global shape of the stability diagram and the position of the boundaries between the different domains converge when averaged over more than 100 simulations. The stochasticity displayed in Fig. 4 results from the random initial perturbations, as described in SI \cite{supplementary}.}, and the colour intensity signifies the frequency at which we find (i) the DIB is stable (green triangle), or it destabilises via (ii) bilayer expansion (red diamond; slightly preferred for polar angle $\lesssim 155^\circ$) or (iii) unzipping (blue square; preferred for polar angle $\gtrsim 155^\circ$) mechanisms upon perturbations.

The stability diagram in Fig. \ref{fig:stable} can be further understood in terms of timescales. The critical parameter is the ratio between lipid adsorption/desorption and lipid exchange timescales. Here we have set $k_{on}/k_{off}=1.6$ and define $\beta_{flow}=\tau_{flow}/\tau_{des}=\Gamma_{\infty} R k_{off}/\xi$. DIBs are stable when $\beta_{flow}$ is large. As detailed in SI \cite{supplementary}, in the stable regime, the system can re-equilibrate without activating lipid flow to the bilayer. In contrast, when the DIBs become unstable for small $\beta_{flow}$, lipid exchange between the monolayers and bilayer is significant. We observe (detailed in SI \cite{supplementary}) the bilayer expansion mechanism is typically accompanied by net desorption and monolayer to bilayer lipid transfer, while for the unzipping mechanism we have net adsorption and bilayer to monolayer lipid transfer.

To conclude, we have studied the stability of DIBs and the dynamic morphologies they follow when they become unstable. Our theory captures the various dynamic pathways observed in experiments. The ``bilayer expansion'' mode is dominant when the timescales for lipid desorptions both in the mono- and bi-layers are slow compared to the timescale for droplet evaporation. However, if one of the desorption timescales is fast, either in the monolayers or bilayer, then the ``unzipping'' mode is preferred. We also predict the presence of crossover behaviours at the boundary between these two dominant modes in the dynamic morphology diagram. Interestingly some of the timescales identified in our theory are size-dependant. This proves to be important for the stability of DIBs, where the timescale for lipid exchange between the mono- and bi-layers becomes faster for smaller DIBs. This drives instability, and for the experimental systems to which we fitted our model, we predict there is a critical size of order 10 $\mu$m to 1 mm below which DIBs can become unstable. An important future work is to exploit this solid theoretical foundation for improving the stability and reproducibility of the DIB platform, which remains a major experimental issue to date, including understanding its possible limit in miniaturisation. Another interesting avenue for research is to complement our model with molecular studies on the lipid transfer mechanisms, which determine the rate constants in our model, and consequently the observed dynamic behaviours.

\begin{acknowledgments}
We thank C. Bain, O. Ces, M. Friddin, Y. Elani, N. Barlow and G. Bolognesi for useful discussions. We acknowledge funding from EPSRC (EP/J017566/1) and Soft Matter and Functional Interfaces Centre for Doctoral Training.
\end{acknowledgments}

\bibliography{biblio}

\begin{thebibliography}{57}
\expandafter\ifx\csname natexlab\endcsname\relax\def\natexlab#1{#1}\fi
\expandafter\ifx\csname bibnamefont\endcsname\relax
  \def\bibnamefont#1{#1}\fi
\expandafter\ifx\csname bibfnamefont\endcsname\relax
  \def\bibfnamefont#1{#1}\fi
\expandafter\ifx\csname citenamefont\endcsname\relax
  \def\citenamefont#1{#1}\fi
\expandafter\ifx\csname url\endcsname\relax
  \def\url#1{\texttt{#1}}\fi
\expandafter\ifx\csname urlprefix\endcsname\relax\def\urlprefix{URL }\fi
\providecommand{\bibinfo}[2]{#2}
\providecommand{\eprint}[2][]{\url{#2}}

\bibitem[{\citenamefont{Funakoshi et~al.}(2006)\citenamefont{Funakoshi, Suzuki,
  and Takeuchi}}]{Funakoshi2006}
\bibinfo{author}{\bibfnamefont{K.}~\bibnamefont{Funakoshi}},
  \bibinfo{author}{\bibfnamefont{H.}~\bibnamefont{Suzuki}}, \bibnamefont{and}
  \bibinfo{author}{\bibfnamefont{S.}~\bibnamefont{Takeuchi}},
  \bibinfo{journal}{Anal. Chem.} \textbf{\bibinfo{volume}{78}},
  \bibinfo{pages}{8169} (\bibinfo{year}{2006}).

\bibitem[{\citenamefont{Bayley et~al.}(2008)\citenamefont{Bayley, Cronin,
  Heron, Holden, Hwang, Syeda, Thompson, and Wallace}}]{Bayley2008DIB}
\bibinfo{author}{\bibfnamefont{H.}~\bibnamefont{Bayley}},
  \bibinfo{author}{\bibfnamefont{B.}~\bibnamefont{Cronin}},
  \bibinfo{author}{\bibfnamefont{A.}~\bibnamefont{Heron}},
  \bibinfo{author}{\bibfnamefont{M.~A.} \bibnamefont{Holden}},
  \bibinfo{author}{\bibfnamefont{W.~L.} \bibnamefont{Hwang}},
  \bibinfo{author}{\bibfnamefont{R.}~\bibnamefont{Syeda}},
  \bibinfo{author}{\bibfnamefont{J.}~\bibnamefont{Thompson}}, \bibnamefont{and}
  \bibinfo{author}{\bibfnamefont{M.}~\bibnamefont{Wallace}},
  \bibinfo{journal}{Mol. BioSyst.} \textbf{\bibinfo{volume}{4}},
  \bibinfo{pages}{1191} (\bibinfo{year}{2008}).

\bibitem[{\citenamefont{Leptihn et~al.}(2013)\citenamefont{Leptihn, Castell,
  Cronin, Lee, Gross, Marshall, Thompson, Holden, and Wallace}}]{Leptihn2013}
\bibinfo{author}{\bibfnamefont{S.}~\bibnamefont{Leptihn}},
  \bibinfo{author}{\bibfnamefont{O.~K.} \bibnamefont{Castell}},
  \bibinfo{author}{\bibfnamefont{B.}~\bibnamefont{Cronin}},
  \bibinfo{author}{\bibfnamefont{E.-H.} \bibnamefont{Lee}},
  \bibinfo{author}{\bibfnamefont{L.~C.~M.} \bibnamefont{Gross}},
  \bibinfo{author}{\bibfnamefont{D.~P.} \bibnamefont{Marshall}},
  \bibinfo{author}{\bibfnamefont{J.~R.} \bibnamefont{Thompson}},
  \bibinfo{author}{\bibfnamefont{M.}~\bibnamefont{Holden}}, \bibnamefont{and}
  \bibinfo{author}{\bibfnamefont{M.~I.} \bibnamefont{Wallace}},
  \bibinfo{journal}{Nat. Protocols} \textbf{\bibinfo{volume}{8}},
  \bibinfo{pages}{1048–1057} (\bibinfo{year}{2013}).

\bibitem[{\citenamefont{Dixit et~al.}(2012)\citenamefont{Dixit, Pincus, Guo,
  and Faris}}]{Dixit2012}
\bibinfo{author}{\bibfnamefont{S.~S.} \bibnamefont{Dixit}},
  \bibinfo{author}{\bibfnamefont{A.}~\bibnamefont{Pincus}},
  \bibinfo{author}{\bibfnamefont{B.}~\bibnamefont{Guo}}, \bibnamefont{and}
  \bibinfo{author}{\bibfnamefont{G.~W.} \bibnamefont{Faris}},
  \bibinfo{journal}{Langmuir} \textbf{\bibinfo{volume}{28}},
  \bibinfo{pages}{7442} (\bibinfo{year}{2012}).

\bibitem[{\citenamefont{Elani et~al.}(2012)\citenamefont{Elani, deMello, Niu,
  and Ces}}]{Elani2012}
\bibinfo{author}{\bibfnamefont{Y.}~\bibnamefont{Elani}},
  \bibinfo{author}{\bibfnamefont{A.~J.} \bibnamefont{deMello}},
  \bibinfo{author}{\bibfnamefont{X.}~\bibnamefont{Niu}}, \bibnamefont{and}
  \bibinfo{author}{\bibfnamefont{O.}~\bibnamefont{Ces}}, \bibinfo{journal}{Lab
  Chip} \textbf{\bibinfo{volume}{12}}, \bibinfo{pages}{3514}
  (\bibinfo{year}{2012}).

\bibitem[{\citenamefont{Thutupalli et~al.}(2013)\citenamefont{Thutupalli,
  Fleury, Steinberger, Herminghaus, and Seemann}}]{Thutupalli2013Micro}
\bibinfo{author}{\bibfnamefont{S.}~\bibnamefont{Thutupalli}},
  \bibinfo{author}{\bibfnamefont{J.-B.} \bibnamefont{Fleury}},
  \bibinfo{author}{\bibfnamefont{A.}~\bibnamefont{Steinberger}},
  \bibinfo{author}{\bibfnamefont{S.}~\bibnamefont{Herminghaus}},
  \bibnamefont{and} \bibinfo{author}{\bibfnamefont{R.}~\bibnamefont{Seemann}},
  \bibinfo{journal}{Chem. Commun.} \textbf{\bibinfo{volume}{49}},
  \bibinfo{pages}{1443} (\bibinfo{year}{2013}).

\bibitem[{\citenamefont{Thiam et~al.}(2012)\citenamefont{Thiam, Bremond, and
  Bibette}}]{Thiam2012}
\bibinfo{author}{\bibfnamefont{A.~R.} \bibnamefont{Thiam}},
  \bibinfo{author}{\bibfnamefont{N.}~\bibnamefont{Bremond}}, \bibnamefont{and}
  \bibinfo{author}{\bibfnamefont{J.}~\bibnamefont{Bibette}},
  \bibinfo{journal}{Langmuir} \textbf{\bibinfo{volume}{28}},
  \bibinfo{pages}{6291} (\bibinfo{year}{2012}).

\bibitem[{\citenamefont{Czekalska et~al.}(2015)\citenamefont{Czekalska,
  Kaminski, Jakiela, Tanuj~Sapra, Bayley, and Garstecki}}]{Czekalska_micro}
\bibinfo{author}{\bibfnamefont{M.~A.} \bibnamefont{Czekalska}},
  \bibinfo{author}{\bibfnamefont{T.~S.} \bibnamefont{Kaminski}},
  \bibinfo{author}{\bibfnamefont{S.}~\bibnamefont{Jakiela}},
  \bibinfo{author}{\bibfnamefont{K.}~\bibnamefont{Tanuj~Sapra}},
  \bibinfo{author}{\bibfnamefont{H.}~\bibnamefont{Bayley}}, \bibnamefont{and}
  \bibinfo{author}{\bibfnamefont{P.}~\bibnamefont{Garstecki}},
  \bibinfo{journal}{Lab Chip} \textbf{\bibinfo{volume}{15}},
  \bibinfo{pages}{541} (\bibinfo{year}{2015}).

\bibitem[{\citenamefont{Villar et~al.}(2013)\citenamefont{Villar, Graham, and
  Bayley}}]{Villar2014Tissue}
\bibinfo{author}{\bibfnamefont{G.}~\bibnamefont{Villar}},
  \bibinfo{author}{\bibfnamefont{A.~D.} \bibnamefont{Graham}},
  \bibnamefont{and} \bibinfo{author}{\bibfnamefont{H.}~\bibnamefont{Bayley}},
  \bibinfo{journal}{Science} \textbf{\bibinfo{volume}{340}},
  \bibinfo{pages}{48} (\bibinfo{year}{2013}).

\bibitem[{\citenamefont{Hwang et~al.}(2007)\citenamefont{Hwang, Holden, White,
  and Bayley}}]{Hwang2007Electrical}
\bibinfo{author}{\bibfnamefont{W.~L.} \bibnamefont{Hwang}},
  \bibinfo{author}{\bibfnamefont{M.~A.} \bibnamefont{Holden}},
  \bibinfo{author}{\bibfnamefont{S.}~\bibnamefont{White}}, \bibnamefont{and}
  \bibinfo{author}{\bibfnamefont{H.}~\bibnamefont{Bayley}},
  \bibinfo{journal}{J. Am. Chem. Soc.} \textbf{\bibinfo{volume}{129}},
  \bibinfo{pages}{11854} (\bibinfo{year}{2007}).

\bibitem[{\citenamefont{Poulos et~al.}(2009{\natexlab{a}})\citenamefont{Poulos,
  Nelson, Jeon, Kim, and Schmidt}}]{Poulos2009}
\bibinfo{author}{\bibfnamefont{J.~L.} \bibnamefont{Poulos}},
  \bibinfo{author}{\bibfnamefont{W.~C.} \bibnamefont{Nelson}},
  \bibinfo{author}{\bibfnamefont{T.-J.} \bibnamefont{Jeon}},
  \bibinfo{author}{\bibfnamefont{C.-J.~â.} \bibnamefont{Kim}},
  \bibnamefont{and} \bibinfo{author}{\bibfnamefont{J.~J.}
  \bibnamefont{Schmidt}}, \bibinfo{journal}{Appl. Phys. Lett.}
  \textbf{\bibinfo{volume}{95}}, \bibinfo{eid}{013706}
  (\bibinfo{year}{2009}{\natexlab{a}}).

\bibitem[{\citenamefont{Freeman et~al.}(2015)\citenamefont{Freeman, Farimani,
  Aluru, and Philen}}]{Freeman2015}
\bibinfo{author}{\bibfnamefont{E.~C.} \bibnamefont{Freeman}},
  \bibinfo{author}{\bibfnamefont{A.~B.} \bibnamefont{Farimani}},
  \bibinfo{author}{\bibfnamefont{N.~R.} \bibnamefont{Aluru}}, \bibnamefont{and}
  \bibinfo{author}{\bibfnamefont{M.~K.} \bibnamefont{Philen}},
  \bibinfo{journal}{Biomicrofluidics} \textbf{\bibinfo{volume}{9}},
  \bibinfo{pages}{064101} (\bibinfo{year}{2015}).

\bibitem[{\citenamefont{Hwang et~al.}(2008)\citenamefont{Hwang, Chen, Cronin,
  Holden, and Bayley}}]{Hwang2008Asymmetric}
\bibinfo{author}{\bibfnamefont{W.~L.} \bibnamefont{Hwang}},
  \bibinfo{author}{\bibfnamefont{M.}~\bibnamefont{Chen}},
  \bibinfo{author}{\bibfnamefont{B.}~\bibnamefont{Cronin}},
  \bibinfo{author}{\bibfnamefont{M.~A.} \bibnamefont{Holden}},
  \bibnamefont{and} \bibinfo{author}{\bibfnamefont{H.}~\bibnamefont{Bayley}},
  \bibinfo{journal}{J. Am. Chem. Soc.} \textbf{\bibinfo{volume}{130}},
  \bibinfo{pages}{5878} (\bibinfo{year}{2008}).

\bibitem[{\citenamefont{Milianta
  et~al.}(2015{\natexlab{a}})\citenamefont{Milianta, Muzzio, Denver, Cawley,
  and Lee}}]{Milianta2015}
\bibinfo{author}{\bibfnamefont{P.~J.} \bibnamefont{Milianta}},
  \bibinfo{author}{\bibfnamefont{M.}~\bibnamefont{Muzzio}},
  \bibinfo{author}{\bibfnamefont{J.}~\bibnamefont{Denver}},
  \bibinfo{author}{\bibfnamefont{G.}~\bibnamefont{Cawley}}, \bibnamefont{and}
  \bibinfo{author}{\bibfnamefont{S.}~\bibnamefont{Lee}},
  \bibinfo{journal}{Langmuir} \textbf{\bibinfo{volume}{31}},
  \bibinfo{pages}{12187} (\bibinfo{year}{2015}{\natexlab{a}}).

\bibitem[{\citenamefont{Elani et~al.}(2013)\citenamefont{Elani, Gee, Law, and
  Ces}}]{Elani2013}
\bibinfo{author}{\bibfnamefont{Y.}~\bibnamefont{Elani}},
  \bibinfo{author}{\bibfnamefont{A.}~\bibnamefont{Gee}},
  \bibinfo{author}{\bibfnamefont{R.~V.} \bibnamefont{Law}}, \bibnamefont{and}
  \bibinfo{author}{\bibfnamefont{O.}~\bibnamefont{Ces}},
  \bibinfo{journal}{Chem. Sci.} \textbf{\bibinfo{volume}{4}},
  \bibinfo{pages}{3332} (\bibinfo{year}{2013}).

\bibitem[{\citenamefont{Schlicht and Zagnoni}(2015)}]{Schlicht2015}
\bibinfo{author}{\bibfnamefont{B.}~\bibnamefont{Schlicht}} \bibnamefont{and}
  \bibinfo{author}{\bibfnamefont{M.}~\bibnamefont{Zagnoni}},
  \bibinfo{journal}{Sci. Rep.} \textbf{\bibinfo{volume}{5}},
  \bibinfo{pages}{9951} (\bibinfo{year}{2015}).

\bibitem[{\citenamefont{Holden et~al.}(2007)\citenamefont{Holden, Needham, and
  Bayley}}]{Holden2007}
\bibinfo{author}{\bibfnamefont{M.~A.} \bibnamefont{Holden}},
  \bibinfo{author}{\bibfnamefont{D.}~\bibnamefont{Needham}}, \bibnamefont{and}
  \bibinfo{author}{\bibfnamefont{H.}~\bibnamefont{Bayley}},
  \bibinfo{journal}{J. Am. Chem. Soc.} \textbf{\bibinfo{volume}{129}},
  \bibinfo{pages}{8650} (\bibinfo{year}{2007}).

\bibitem[{\citenamefont{Barriga et~al.}(2014)\citenamefont{Barriga, Booth,
  Haylock, Bazin, Templer, and Ces}}]{Barriga2014}
\bibinfo{author}{\bibfnamefont{H.~M.~G.} \bibnamefont{Barriga}},
  \bibinfo{author}{\bibfnamefont{P.}~\bibnamefont{Booth}},
  \bibinfo{author}{\bibfnamefont{S.}~\bibnamefont{Haylock}},
  \bibinfo{author}{\bibfnamefont{R.}~\bibnamefont{Bazin}},
  \bibinfo{author}{\bibfnamefont{R.~H.} \bibnamefont{Templer}},
  \bibnamefont{and} \bibinfo{author}{\bibfnamefont{O.}~\bibnamefont{Ces}},
  \bibinfo{journal}{J. R. Soc. Interface} \textbf{\bibinfo{volume}{11}},
  \bibinfo{pages}{20140404} (\bibinfo{year}{2014}).

\bibitem[{\citenamefont{Syeda et~al.}(2008)\citenamefont{Syeda, Holden, Hwang,
  and Bayley}}]{Syeda2008}
\bibinfo{author}{\bibfnamefont{R.}~\bibnamefont{Syeda}},
  \bibinfo{author}{\bibfnamefont{M.~A.} \bibnamefont{Holden}},
  \bibinfo{author}{\bibfnamefont{W.~L.} \bibnamefont{Hwang}}, \bibnamefont{and}
  \bibinfo{author}{\bibfnamefont{H.}~\bibnamefont{Bayley}},
  \bibinfo{journal}{J. Am. Chem. Soc.} \textbf{\bibinfo{volume}{130}},
  \bibinfo{pages}{15543} (\bibinfo{year}{2008}).

\bibitem[{\citenamefont{Poulos et~al.}(2009{\natexlab{b}})\citenamefont{Poulos,
  Jeon, Damoiseaux, Gillespie, Bradley, and Schmidt}}]{Poulos2009Ion}
\bibinfo{author}{\bibfnamefont{J.~L.} \bibnamefont{Poulos}},
  \bibinfo{author}{\bibfnamefont{T.-J.} \bibnamefont{Jeon}},
  \bibinfo{author}{\bibfnamefont{R.}~\bibnamefont{Damoiseaux}},
  \bibinfo{author}{\bibfnamefont{E.~J.} \bibnamefont{Gillespie}},
  \bibinfo{author}{\bibfnamefont{K.~A.} \bibnamefont{Bradley}},
  \bibnamefont{and} \bibinfo{author}{\bibfnamefont{J.~J.}
  \bibnamefont{Schmidt}}, \bibinfo{journal}{Biosens. Bioelectron.}
  \textbf{\bibinfo{volume}{24}}, \bibinfo{pages}{1806 }
  (\bibinfo{year}{2009}{\natexlab{b}}).

\bibitem[{\citenamefont{Elani et~al.}(2014)\citenamefont{Elani, Law, and
  Ces}}]{Elani2014}
\bibinfo{author}{\bibfnamefont{Y.}~\bibnamefont{Elani}},
  \bibinfo{author}{\bibfnamefont{R.~V.} \bibnamefont{Law}}, \bibnamefont{and}
  \bibinfo{author}{\bibfnamefont{O.}~\bibnamefont{Ces}}, \bibinfo{journal}{Nat.
  Commun.} \textbf{\bibinfo{volume}{5}}, \bibinfo{pages}{5305}
  (\bibinfo{year}{2014}).

\bibitem[{\citenamefont{Elani et~al.}(2016)\citenamefont{Elani, Solvas, Edel,
  Law, and Ces}}]{Elani2016Microfluidic}
\bibinfo{author}{\bibfnamefont{Y.}~\bibnamefont{Elani}},
  \bibinfo{author}{\bibfnamefont{X.~C.~I.} \bibnamefont{Solvas}},
  \bibinfo{author}{\bibfnamefont{J.~B.} \bibnamefont{Edel}},
  \bibinfo{author}{\bibfnamefont{R.~V.} \bibnamefont{Law}}, \bibnamefont{and}
  \bibinfo{author}{\bibfnamefont{O.}~\bibnamefont{Ces}},
  \bibinfo{journal}{Chem. Commun.} \textbf{\bibinfo{volume}{52}},
  \bibinfo{pages}{5961} (\bibinfo{year}{2016}).

\bibitem[{\citenamefont{Punnamaraju et~al.}(2012)\citenamefont{Punnamaraju,
  You, and Steckl}}]{Punnamaraju2012}
\bibinfo{author}{\bibfnamefont{S.}~\bibnamefont{Punnamaraju}},
  \bibinfo{author}{\bibfnamefont{H.}~\bibnamefont{You}}, \bibnamefont{and}
  \bibinfo{author}{\bibfnamefont{A.~J.} \bibnamefont{Steckl}},
  \bibinfo{journal}{Langmuir} \textbf{\bibinfo{volume}{28}},
  \bibinfo{pages}{7657} (\bibinfo{year}{2012}).

\bibitem[{\citenamefont{Zhang et~al.}(2016)\citenamefont{Zhang, Wan, Schwarz,
  and Bowick}}]{Zhang2016Shape}
\bibinfo{author}{\bibfnamefont{T.}~\bibnamefont{Zhang}},
  \bibinfo{author}{\bibfnamefont{D.}~\bibnamefont{Wan}},
  \bibinfo{author}{\bibfnamefont{J.~M.} \bibnamefont{Schwarz}},
  \bibnamefont{and} \bibinfo{author}{\bibfnamefont{M.~J.}
  \bibnamefont{Bowick}}, \bibinfo{journal}{Phys. Rev. Lett.}
  \textbf{\bibinfo{volume}{116}}, \bibinfo{pages}{108301}
  (\bibinfo{year}{2016}).

\bibitem[{\citenamefont{Maglia et~al.}(2009)\citenamefont{Maglia, Heron, Hwang,
  Holden, Mikhailova, Li, Cheley, and Bayley}}]{Maglia2009Droplet}
\bibinfo{author}{\bibfnamefont{G.}~\bibnamefont{Maglia}},
  \bibinfo{author}{\bibfnamefont{A.~J.} \bibnamefont{Heron}},
  \bibinfo{author}{\bibfnamefont{W.~L.} \bibnamefont{Hwang}},
  \bibinfo{author}{\bibfnamefont{M.~A.} \bibnamefont{Holden}},
  \bibinfo{author}{\bibfnamefont{E.}~\bibnamefont{Mikhailova}},
  \bibinfo{author}{\bibfnamefont{Q.}~\bibnamefont{Li}},
  \bibinfo{author}{\bibfnamefont{S.}~\bibnamefont{Cheley}}, \bibnamefont{and}
  \bibinfo{author}{\bibfnamefont{H.}~\bibnamefont{Bayley}},
  \bibinfo{journal}{Nat. Nano} \textbf{\bibinfo{volume}{4}},
  \bibinfo{pages}{437–440} (\bibinfo{year}{2009}).

\bibitem[{\citenamefont{Boreyko et~al.}(2013)\citenamefont{Boreyko,
  Mruetusatorn, Sarles, Retterer, and Collier}}]{boreyko2013evaporation}
\bibinfo{author}{\bibfnamefont{J.~B.} \bibnamefont{Boreyko}},
  \bibinfo{author}{\bibfnamefont{P.}~\bibnamefont{Mruetusatorn}},
  \bibinfo{author}{\bibfnamefont{S.~A.} \bibnamefont{Sarles}},
  \bibinfo{author}{\bibfnamefont{S.~T.} \bibnamefont{Retterer}},
  \bibnamefont{and} \bibinfo{author}{\bibfnamefont{C.~P.}
  \bibnamefont{Collier}}, \bibinfo{journal}{J. Am. Chem. Soc.}
  \textbf{\bibinfo{volume}{135}}, \bibinfo{pages}{5545} (\bibinfo{year}{2013}).

\bibitem[{\citenamefont{Mruetusatorn et~al.}(2014)\citenamefont{Mruetusatorn,
  Boreyko, Venkatesan, Sarles, Hayes, and Collier}}]{mruetusatorn2014dynamic}
\bibinfo{author}{\bibfnamefont{P.}~\bibnamefont{Mruetusatorn}},
  \bibinfo{author}{\bibfnamefont{J.~B.} \bibnamefont{Boreyko}},
  \bibinfo{author}{\bibfnamefont{G.~A.} \bibnamefont{Venkatesan}},
  \bibinfo{author}{\bibfnamefont{S.~A.} \bibnamefont{Sarles}},
  \bibinfo{author}{\bibfnamefont{D.~G.} \bibnamefont{Hayes}}, \bibnamefont{and}
  \bibinfo{author}{\bibfnamefont{C.~P.} \bibnamefont{Collier}},
  \bibinfo{journal}{Soft Matter} \textbf{\bibinfo{volume}{10}},
  \bibinfo{pages}{2530} (\bibinfo{year}{2014}).

\bibitem[{\citenamefont{Eastoe and Dalton}(2000)}]{eastoe2000dynamic}
\bibinfo{author}{\bibfnamefont{J.}~\bibnamefont{Eastoe}} \bibnamefont{and}
  \bibinfo{author}{\bibfnamefont{J.}~\bibnamefont{Dalton}},
  \bibinfo{journal}{Adv. Colloid Interface Sci.} \textbf{\bibinfo{volume}{85}},
  \bibinfo{pages}{103} (\bibinfo{year}{2000}).

\bibitem[{\citenamefont{Venkatesan et~al.}(2015)\citenamefont{Venkatesan, Lee,
  Farimani, Heiranian, Collier, Aluru, and Sarles}}]{venkatesan2015adsorption}
\bibinfo{author}{\bibfnamefont{G.~A.} \bibnamefont{Venkatesan}},
  \bibinfo{author}{\bibfnamefont{J.}~\bibnamefont{Lee}},
  \bibinfo{author}{\bibfnamefont{A.~B.} \bibnamefont{Farimani}},
  \bibinfo{author}{\bibfnamefont{M.}~\bibnamefont{Heiranian}},
  \bibinfo{author}{\bibfnamefont{C.~P.} \bibnamefont{Collier}},
  \bibinfo{author}{\bibfnamefont{N.~R.} \bibnamefont{Aluru}}, \bibnamefont{and}
  \bibinfo{author}{\bibfnamefont{S.~A.} \bibnamefont{Sarles}},
  \bibinfo{journal}{Langmuir} \textbf{\bibinfo{volume}{31}},
  \bibinfo{pages}{12883} (\bibinfo{year}{2015}).

\bibitem[{\citenamefont{Doi}(2013)}]{Doi2013}
\bibinfo{author}{\bibfnamefont{M.}~\bibnamefont{Doi}},
  \emph{\bibinfo{title}{Soft Matter Physics}} (\bibinfo{publisher}{Oxford
  University Press}, \bibinfo{address}{Oxford}, \bibinfo{year}{2013}).

\bibitem[{\citenamefont{Liu and Messow}(2000)}]{liu2000diffusion}
\bibinfo{author}{\bibfnamefont{J.}~\bibnamefont{Liu}} \bibnamefont{and}
  \bibinfo{author}{\bibfnamefont{U.}~\bibnamefont{Messow}},
  \bibinfo{journal}{Colloid Polym. Sci.} \textbf{\bibinfo{volume}{278}},
  \bibinfo{pages}{124} (\bibinfo{year}{2000}).

\bibitem[{sup()}]{supplementary}
\bibinfo{note}{See Supplemental Material at \url{url} for additional
  discussions on the lipid kinetics model for the lipid bilayer, the Markov
  Chain Monte Carlo methods, observed crossover behaviours between ``bilayer
  expansion'' and ``unzipping'' modes, comparison of best-fit parameter values
  against typical literature values, and further analysis on DIB stability. The
  Supplemental Material includes Refs.
  \cite{schutz1997single,filippov2003effect,derzko1980comparative,gassin2013surfactant,Bennett2009,Grafmuller2013,menichetti2017efficient,Montel2011,Polenz2015,Sleeboom2017,zhou2013facile,kirby2015sequential,he2015adsorption,bleys1985adsorption,Ferri200813,lis1982measurement,israelachvili2011intermolecular}.}

\bibitem[{\citenamefont{Venable et~al.}(2015)\citenamefont{Venable, Brown, and
  Pastor}}]{venable2015mechanical}
\bibinfo{author}{\bibfnamefont{R.~M.} \bibnamefont{Venable}},
  \bibinfo{author}{\bibfnamefont{F.~L.} \bibnamefont{Brown}}, \bibnamefont{and}
  \bibinfo{author}{\bibfnamefont{R.~W.} \bibnamefont{Pastor}},
  \bibinfo{journal}{Chem. Phys. Lipids} \textbf{\bibinfo{volume}{192}},
  \bibinfo{pages}{60} (\bibinfo{year}{2015}).

\bibitem[{\citenamefont{Staykova et~al.}(2013)\citenamefont{Staykova, Arroyo,
  Rahimi, and Stone}}]{staykova2013confined}
\bibinfo{author}{\bibfnamefont{M.}~\bibnamefont{Staykova}},
  \bibinfo{author}{\bibfnamefont{M.}~\bibnamefont{Arroyo}},
  \bibinfo{author}{\bibfnamefont{M.}~\bibnamefont{Rahimi}}, \bibnamefont{and}
  \bibinfo{author}{\bibfnamefont{H.~A.} \bibnamefont{Stone}},
  \bibinfo{journal}{Phys. Rev. Lett.} \textbf{\bibinfo{volume}{110}},
  \bibinfo{pages}{028101} (\bibinfo{year}{2013}).

\bibitem[{\citenamefont{Walter and Gutknecht}(1986)}]{walter1986permeability}
\bibinfo{author}{\bibfnamefont{A.}~\bibnamefont{Walter}} \bibnamefont{and}
  \bibinfo{author}{\bibfnamefont{J.}~\bibnamefont{Gutknecht}},
  \bibinfo{journal}{J. Membr. Biol.} \textbf{\bibinfo{volume}{90}},
  \bibinfo{pages}{207} (\bibinfo{year}{1986}).

\bibitem[{\citenamefont{Mathai et~al.}(2008)\citenamefont{Mathai,
  Tristram-Nagle, Nagle, and Zeidel}}]{mathai2008structural}
\bibinfo{author}{\bibfnamefont{J.~C.} \bibnamefont{Mathai}},
  \bibinfo{author}{\bibfnamefont{S.}~\bibnamefont{Tristram-Nagle}},
  \bibinfo{author}{\bibfnamefont{J.~F.} \bibnamefont{Nagle}}, \bibnamefont{and}
  \bibinfo{author}{\bibfnamefont{M.~L.} \bibnamefont{Zeidel}},
  \bibinfo{journal}{J. Gen. Physiol.} \textbf{\bibinfo{volume}{131}},
  \bibinfo{pages}{69} (\bibinfo{year}{2008}).

\bibitem[{\citenamefont{Milianta
  et~al.}(2015{\natexlab{b}})\citenamefont{Milianta, Muzzio, Denver, Cawley,
  and Lee}}]{milianta2015water}
\bibinfo{author}{\bibfnamefont{P.~J.} \bibnamefont{Milianta}},
  \bibinfo{author}{\bibfnamefont{M.}~\bibnamefont{Muzzio}},
  \bibinfo{author}{\bibfnamefont{J.}~\bibnamefont{Denver}},
  \bibinfo{author}{\bibfnamefont{G.}~\bibnamefont{Cawley}}, \bibnamefont{and}
  \bibinfo{author}{\bibfnamefont{S.}~\bibnamefont{Lee}},
  \bibinfo{journal}{Langmuir} \textbf{\bibinfo{volume}{31}},
  \bibinfo{pages}{12187} (\bibinfo{year}{2015}{\natexlab{b}}).

\bibitem[{\citenamefont{Olbrich et~al.}(2000)\citenamefont{Olbrich, Rawicz,
  Needham, and Evans}}]{olbrich2000water}
\bibinfo{author}{\bibfnamefont{K.}~\bibnamefont{Olbrich}},
  \bibinfo{author}{\bibfnamefont{W.}~\bibnamefont{Rawicz}},
  \bibinfo{author}{\bibfnamefont{D.}~\bibnamefont{Needham}}, \bibnamefont{and}
  \bibinfo{author}{\bibfnamefont{E.}~\bibnamefont{Evans}},
  \bibinfo{journal}{Biophys. J.} \textbf{\bibinfo{volume}{79}},
  \bibinfo{pages}{321} (\bibinfo{year}{2000}).

\bibitem[{\citenamefont{Finkelstein}(1976)}]{finkelstein1976water}
\bibinfo{author}{\bibfnamefont{A.}~\bibnamefont{Finkelstein}},
  \bibinfo{journal}{J. Gen. Physiol.} \textbf{\bibinfo{volume}{68}},
  \bibinfo{pages}{127} (\bibinfo{year}{1976}).

\bibitem[{\citenamefont{Sehgal et~al.}(2003)\citenamefont{Sehgal, Doe, and
  Sharma}}]{sehgal2003micellar}
\bibinfo{author}{\bibfnamefont{P.}~\bibnamefont{Sehgal}},
  \bibinfo{author}{\bibfnamefont{H.}~\bibnamefont{Doe}}, \bibnamefont{and}
  \bibinfo{author}{\bibfnamefont{M.}~\bibnamefont{Sharma}},
  \bibinfo{journal}{Colloid Polym. Sci.} \textbf{\bibinfo{volume}{282}},
  \bibinfo{pages}{188} (\bibinfo{year}{2003}).

\bibitem[{\citenamefont{Zhou et~al.}(2013)\citenamefont{Zhou, Yao, Chen, Li,
  and Yao}}]{zhou2013facile}
\bibinfo{author}{\bibfnamefont{H.}~\bibnamefont{Zhou}},
  \bibinfo{author}{\bibfnamefont{Y.}~\bibnamefont{Yao}},
  \bibinfo{author}{\bibfnamefont{Q.}~\bibnamefont{Chen}},
  \bibinfo{author}{\bibfnamefont{G.}~\bibnamefont{Li}}, \bibnamefont{and}
  \bibinfo{author}{\bibfnamefont{S.}~\bibnamefont{Yao}},
  \bibinfo{journal}{Appl. Phys. Lett.} \textbf{\bibinfo{volume}{103}},
  \bibinfo{pages}{234102} (\bibinfo{year}{2013}).

\bibitem[{\citenamefont{Sch{\"u}tz et~al.}(1997)\citenamefont{Sch{\"u}tz,
  Schindler, and Schmidt}}]{schutz1997single}
\bibinfo{author}{\bibfnamefont{G.~J.} \bibnamefont{Sch{\"u}tz}},
  \bibinfo{author}{\bibfnamefont{H.}~\bibnamefont{Schindler}},
  \bibnamefont{and} \bibinfo{author}{\bibfnamefont{T.}~\bibnamefont{Schmidt}},
  \bibinfo{journal}{Biophysical journal} \textbf{\bibinfo{volume}{73}},
  \bibinfo{pages}{1073} (\bibinfo{year}{1997}).

\bibitem[{\citenamefont{Filippov et~al.}(2003)\citenamefont{Filippov,
  Or{\"a}dd, and Lindblom}}]{filippov2003effect}
\bibinfo{author}{\bibfnamefont{A.}~\bibnamefont{Filippov}},
  \bibinfo{author}{\bibfnamefont{G.}~\bibnamefont{Or{\"a}dd}},
  \bibnamefont{and} \bibinfo{author}{\bibfnamefont{G.}~\bibnamefont{Lindblom}},
  \bibinfo{journal}{Biophysical journal} \textbf{\bibinfo{volume}{84}},
  \bibinfo{pages}{3079} (\bibinfo{year}{2003}).

\bibitem[{\citenamefont{Derzko and Jacobson}(1980)}]{derzko1980comparative}
\bibinfo{author}{\bibfnamefont{Z.}~\bibnamefont{Derzko}} \bibnamefont{and}
  \bibinfo{author}{\bibfnamefont{K.}~\bibnamefont{Jacobson}},
  \bibinfo{journal}{Biochemistry} \textbf{\bibinfo{volume}{19}},
  \bibinfo{pages}{6050} (\bibinfo{year}{1980}).

\bibitem[{\citenamefont{Gassin et~al.}(2013)\citenamefont{Gassin, Champory,
  Martin-Gassin, Dufr{\^e}che, and Diat}}]{gassin2013surfactant}
\bibinfo{author}{\bibfnamefont{P.-M.} \bibnamefont{Gassin}},
  \bibinfo{author}{\bibfnamefont{R.}~\bibnamefont{Champory}},
  \bibinfo{author}{\bibfnamefont{G.}~\bibnamefont{Martin-Gassin}},
  \bibinfo{author}{\bibfnamefont{J.-F.} \bibnamefont{Dufr{\^e}che}},
  \bibnamefont{and} \bibinfo{author}{\bibfnamefont{O.}~\bibnamefont{Diat}},
  \bibinfo{journal}{Colloids Surf., A} \textbf{\bibinfo{volume}{436}},
  \bibinfo{pages}{1103} (\bibinfo{year}{2013}).

\bibitem[{\citenamefont{Bennett et~al.}(2009)\citenamefont{Bennett, MacCallum,
  Hinner, Marrink, and Tieleman}}]{Bennett2009}
\bibinfo{author}{\bibfnamefont{W.~F.~D.} \bibnamefont{Bennett}},
  \bibinfo{author}{\bibfnamefont{J.~L.} \bibnamefont{MacCallum}},
  \bibinfo{author}{\bibfnamefont{M.~J.} \bibnamefont{Hinner}},
  \bibinfo{author}{\bibfnamefont{S.~J.} \bibnamefont{Marrink}},
  \bibnamefont{and} \bibinfo{author}{\bibfnamefont{D.~P.}
  \bibnamefont{Tieleman}}, \bibinfo{journal}{J. Am. Chem. Soc.}
  \textbf{\bibinfo{volume}{131}}, \bibinfo{pages}{12714}
  (\bibinfo{year}{2009}).

\bibitem[{\citenamefont{Grafm\"{u}ller
  et~al.}(2013)\citenamefont{Grafm\"{u}ller, Lipowsky, and
  Knecht}}]{Grafmuller2013}
\bibinfo{author}{\bibfnamefont{A.}~\bibnamefont{Grafm\"{u}ller}},
  \bibinfo{author}{\bibfnamefont{R.}~\bibnamefont{Lipowsky}}, \bibnamefont{and}
  \bibinfo{author}{\bibfnamefont{V.}~\bibnamefont{Knecht}},
  \bibinfo{journal}{Phys. Chem. Chem. Phys.} \textbf{\bibinfo{volume}{15}},
  \bibinfo{pages}{876} (\bibinfo{year}{2013}).

\bibitem[{\citenamefont{Menichetti et~al.}(2017)\citenamefont{Menichetti,
  Kremer, and Bereau}}]{menichetti2017efficient}
\bibinfo{author}{\bibfnamefont{R.}~\bibnamefont{Menichetti}},
  \bibinfo{author}{\bibfnamefont{K.}~\bibnamefont{Kremer}}, \bibnamefont{and}
  \bibinfo{author}{\bibfnamefont{T.}~\bibnamefont{Bereau}},
  \bibinfo{journal}{Biochemical and biophysical research communications}
  (\bibinfo{year}{2017}).

\bibitem[{\citenamefont{Montel et~al.}(2011)\citenamefont{Montel, Delarue,
  Elgeti, Malaquin, Basan, Risler, Cabane, Vignjevic, Prost, Cappello
  et~al.}}]{Montel2011}
\bibinfo{author}{\bibfnamefont{F.}~\bibnamefont{Montel}},
  \bibinfo{author}{\bibfnamefont{M.}~\bibnamefont{Delarue}},
  \bibinfo{author}{\bibfnamefont{J.}~\bibnamefont{Elgeti}},
  \bibinfo{author}{\bibfnamefont{L.}~\bibnamefont{Malaquin}},
  \bibinfo{author}{\bibfnamefont{M.}~\bibnamefont{Basan}},
  \bibinfo{author}{\bibfnamefont{T.}~\bibnamefont{Risler}},
  \bibinfo{author}{\bibfnamefont{B.}~\bibnamefont{Cabane}},
  \bibinfo{author}{\bibfnamefont{D.}~\bibnamefont{Vignjevic}},
  \bibinfo{author}{\bibfnamefont{J.}~\bibnamefont{Prost}},
  \bibinfo{author}{\bibfnamefont{G.}~\bibnamefont{Cappello}},
  \bibnamefont{et~al.}, \bibinfo{journal}{Phys. Rev. Lett.}
  \textbf{\bibinfo{volume}{107}}, \bibinfo{pages}{188102}
  (\bibinfo{year}{2011}).

\bibitem[{\citenamefont{Polenz et~al.}(2015)\citenamefont{Polenz, Weitz, and
  Baret}}]{Polenz2015}
\bibinfo{author}{\bibfnamefont{I.}~\bibnamefont{Polenz}},
  \bibinfo{author}{\bibfnamefont{D.~A.} \bibnamefont{Weitz}}, \bibnamefont{and}
  \bibinfo{author}{\bibfnamefont{J.~C.} \bibnamefont{Baret}},
  \bibinfo{journal}{Langmuir} \textbf{\bibinfo{volume}{31}},
  \bibinfo{pages}{1127} (\bibinfo{year}{2015}).

\bibitem[{\citenamefont{Sleeboom et~al.}(2017)\citenamefont{Sleeboom,
  Voudouris, Punter, Aangenendt, Florea, van~der Schoot, and
  Wyss}}]{Sleeboom2017}
\bibinfo{author}{\bibfnamefont{J.~J.~F.} \bibnamefont{Sleeboom}},
  \bibinfo{author}{\bibfnamefont{P.}~\bibnamefont{Voudouris}},
  \bibinfo{author}{\bibfnamefont{M.~T. J. J.~M.} \bibnamefont{Punter}},
  \bibinfo{author}{\bibfnamefont{F.~J.} \bibnamefont{Aangenendt}},
  \bibinfo{author}{\bibfnamefont{D.}~\bibnamefont{Florea}},
  \bibinfo{author}{\bibfnamefont{P.}~\bibnamefont{van~der Schoot}},
  \bibnamefont{and} \bibinfo{author}{\bibfnamefont{H.~M.} \bibnamefont{Wyss}},
  \bibinfo{journal}{Phys. Rev. Lett.} \textbf{\bibinfo{volume}{119}},
  \bibinfo{pages}{098001} (\bibinfo{year}{2017}).

\bibitem[{\citenamefont{Kirby et~al.}(2015)\citenamefont{Kirby, Anna, and
  Walker}}]{kirby2015sequential}
\bibinfo{author}{\bibfnamefont{S.~M.} \bibnamefont{Kirby}},
  \bibinfo{author}{\bibfnamefont{S.~L.} \bibnamefont{Anna}}, \bibnamefont{and}
  \bibinfo{author}{\bibfnamefont{L.~M.} \bibnamefont{Walker}},
  \bibinfo{journal}{Langmuir} \textbf{\bibinfo{volume}{31}},
  \bibinfo{pages}{4063} (\bibinfo{year}{2015}).

\bibitem[{\citenamefont{He et~al.}(2015)\citenamefont{He, Yazhgur, Salonen, and
  Langevin}}]{he2015adsorption}
\bibinfo{author}{\bibfnamefont{Y.}~\bibnamefont{He}},
  \bibinfo{author}{\bibfnamefont{P.}~\bibnamefont{Yazhgur}},
  \bibinfo{author}{\bibfnamefont{A.}~\bibnamefont{Salonen}}, \bibnamefont{and}
  \bibinfo{author}{\bibfnamefont{D.}~\bibnamefont{Langevin}},
  \bibinfo{journal}{Adv. Colloid Interface Sci.}
  \textbf{\bibinfo{volume}{222}}, \bibinfo{pages}{377} (\bibinfo{year}{2015}).

\bibitem[{\citenamefont{Bleys and Joos}(1985)}]{bleys1985adsorption}
\bibinfo{author}{\bibfnamefont{G.}~\bibnamefont{Bleys}} \bibnamefont{and}
  \bibinfo{author}{\bibfnamefont{P.}~\bibnamefont{Joos}}, \bibinfo{journal}{J.
  Phys. Chem.} \textbf{\bibinfo{volume}{89}}, \bibinfo{pages}{1027}
  (\bibinfo{year}{1985}).

\bibitem[{\citenamefont{Ferri et~al.}(2008)\citenamefont{Ferri, Gorevski,
  Kotsmar, Leser, and Miller}}]{Ferri200813}
\bibinfo{author}{\bibfnamefont{J.~K.} \bibnamefont{Ferri}},
  \bibinfo{author}{\bibfnamefont{N.}~\bibnamefont{Gorevski}},
  \bibinfo{author}{\bibfnamefont{C.}~\bibnamefont{Kotsmar}},
  \bibinfo{author}{\bibfnamefont{M.~E.} \bibnamefont{Leser}}, \bibnamefont{and}
  \bibinfo{author}{\bibfnamefont{R.}~\bibnamefont{Miller}},
  \bibinfo{journal}{Colloids Surf., A} \textbf{\bibinfo{volume}{319}},
  \bibinfo{pages}{13 } (\bibinfo{year}{2008}).

\bibitem[{\citenamefont{Lis et~al.}(1982)\citenamefont{Lis, McAlister, Fuller,
  Rand, and Parsegian}}]{lis1982measurement}
\bibinfo{author}{\bibfnamefont{L.}~\bibnamefont{Lis}},
  \bibinfo{author}{\bibfnamefont{M.}~\bibnamefont{McAlister}},
  \bibinfo{author}{\bibfnamefont{N.}~\bibnamefont{Fuller}},
  \bibinfo{author}{\bibfnamefont{R.}~\bibnamefont{Rand}}, \bibnamefont{and}
  \bibinfo{author}{\bibfnamefont{V.}~\bibnamefont{Parsegian}},
  \bibinfo{journal}{Biophysical journal} \textbf{\bibinfo{volume}{37}},
  \bibinfo{pages}{667} (\bibinfo{year}{1982}).

\bibitem[{\citenamefont{Israelachvili}(2011)}]{israelachvili2011intermolecular}
\bibinfo{author}{\bibfnamefont{J.~N.} \bibnamefont{Israelachvili}},
  \emph{\bibinfo{title}{Intermolecular and surface forces}}
  (\bibinfo{publisher}{Academic press}, \bibinfo{year}{2011}).

\end{thebibliography}

\end{document}